\documentclass[aps, prl, twocolumn,amsmath,amssymb,floatfix,superscriptaddress,prl]{revtex4-1}
\usepackage{leading}
\usepackage[latin3]{inputenc}
\usepackage[makeroom]{cancel}
\usepackage{graphicx}
\usepackage{amsmath}
\usepackage{amsfonts}
\usepackage{amssymb}
\usepackage{color}
\usepackage[left=2cm,right=2cm,top=2cm,bottom=2cm]{geometry}
\usepackage{graphicx} 
\usepackage{dcolumn}
\usepackage{bm}
\usepackage{simplewick}
\usepackage{array}
\usepackage{appendix}
\usepackage{cancel}
\usepackage[normalem]{ulem} 
\RequirePackage[
   hyperindex,colorlinks,bookmarksnumbered,
   plainpages=true,pdfstartview=FitH]{hyperref}
\hypersetup{linkcolor=blue,urlcolor=blue,citecolor=blue}
\usepackage{hyperref}
\usepackage{mathrsfs}

 \renewcommand{\emph}[1]{\textit{#1}}

\definecolor{darkblue}{rgb}{0,0,0.5}
\definecolor{darkgreen}{rgb}{0,0.5,0}
\definecolor{darkred}{rgb}{.7,0,0}
\definecolor{purple}{rgb}{0.5,0,0.6}
\definecolor{orange}{rgb}{1,0.5,0}
\definecolor{grey}{rgb}{.6,.6,.6}
\definecolor{lightpink}{rgb}{1,0.7,0.75}
\definecolor{pink}{rgb}{1,0.4,0.58}
\definecolor{deeppink}{rgb}{1,0.08,0.58}
\usepackage{upgreek}

\begin{document}
\title{Coupled quantum critical states in a circuit simulator}
\author{D. B. Karki}
\email{deep.b.karki@gmail.com}
\affiliation{The Abdus Salam International Centre for Theoretical Physics (ICTP), Strada Costiera 11, I-34151 Trieste, Italy}
\author{Andrew K. Mitchell}
\email{Andrew.Mitchell@ucd.ie}
\affiliation{School of Physics, University College Dublin, Belfield, Dublin 4, Ireland}
\affiliation{Rinn Quantum and Centre for Quantum Engineering, Science, and Technology,\\School of Physics, University College Dublin, Dublin 4, Ireland}


\begin{abstract}
Hybrid metal-semiconductor quantum circuits offer a controlled setting for realizing and probing strongly correlated quantum matter. A largely open question that may be addressed by such simulators is whether coupled critical states can combine into new non-Fermi liquid (NFL) phases.  Inspired by the recent realization of a two-site charge-Kondo circuit [W. Pouse \textit{et al.}, \href{https://www.nature.com/articles/s41567-022-01905-4}{Nat. Phys. \textbf{19}, 492 (2023)}], we uncover the fate of coupled Kondo anyons emerging from the two-channel-Kondo critical point at each site. Bosonization and quantum Brownian motion methods yield the full quasiballistic phase diagram and low-temperature transport properties, while numerical renormalization group calculations resolve the complementary weak-coupling limit. Together, these provide a unified description, and reveal robust NFL phases with continuously tunable transport and susceptibility exponents. Stronger inter-site coupling ultimately produces a screened Fermi liquid. 
\end{abstract}

\maketitle


\textit{\textbf{Introduction.--}} Nanoelectronic circuits built from hybrid metal-semiconductor quantum dot components provide a versatile platform for simulating correlated states of quantum matter~\cite{Pierre_2015,Pierre_2018,pouse2021,Pierre_2018a,piquard2023observing,piquard2026experimental,Sri}. These devices comprise micron-scale metallic grains, contacted to gate-defined regions of a semiconductor two-dimensional electron gas in the quantum Hall regime, and coupled to spin-polarized edge channels through tunable quantum point contacts (QPCs). The metallic island supplies a quasi-continuous spectrum, while its finite charging energy quantizes the island charge~\cite{Pierre_2016}. Charge fluctuations through the QPCs then generate strong many-body correlations at low temperatures~\cite{Matveev_1991,Matveev1995}.

Non-Fermi liquid (NFL) behavior can arise in these devices near impurity quantum phase transitions~\cite{Pierre_2015,Pierre_2018,pouse2021} associated with multichannel Kondo effects~\cite{Nozieres_Blandin_JPhys_1980,al1,Mitchell_2012a,karki2020quantum,mitchell2021so,*sen2024many}. In particular, a single island coupled to two leads gives two-channel Kondo (2CK) physics~\cite{Furusaki1995b}, which supports a critical point driven by the frustrated screening of macroscopic charge degrees of freedom on the island.  The experimental realizations in Refs.~\cite{Pierre_2015,Pierre_2018} achieve precise agreement with theoretical predictions for transport from the 2CK model~\cite{Furusaki1995b,Sela_2016}, while the observation of a fractional impurity entropy~\cite{han2022fractional,piquard2026experimental}
indicates fractionalization at the critical point, where a localized Majorana zero mode is predicted to emerge~\cite{EK1}. More exotic critical points in three-lead or coupled-island structures have also been reported~\cite{Pierre_2018,pouse2021}. 
This demonstrates that nontrivial critical states and Kondo anyons~\cite{kam,lopes2020anyons,KBM1} can be engineered and manipulated in such devices.

\begin{figure}[b]
\includegraphics[scale=0.7]{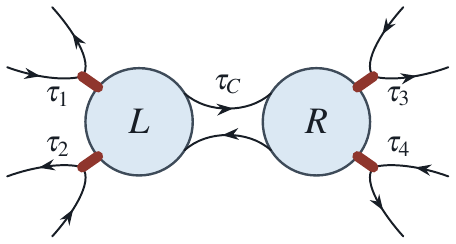}
\caption{Two-site charge-Kondo circuit, in which two coupled metallic islands are each connected to two external leads. The setup allows to explore the fate of coupled critical states.}\label{fig1}
\end{figure}

In addition to multichannel physics, coupling multiple Kondo impurities introduces further competition. In the conventional two-impurity Kondo (2IK) model, separate Kondo screening competes with inter-impurity singlet formation, resulting in a NFL critical point~\cite{a1,alaf} in the 2CK universality class~\cite{Mitchell_2012b}. 
However, the two-site charge-Kondo setup recently realized in Ref.~\cite{pouse2021} offers more intriguing possibilities, due to cooperative charge transfer (and hence Kondo exchange) between islands at the central QPC~\cite{dde1}. In the experimental setup, each island was coupled to two external leads (see Fig.~\ref{fig1}), but the device was operated in a reduced configuration with only one outer channel on each side kept open. That case was studied theoretically in the weak-coupling Kondo limit in Ref.~\cite{KBM1} and the complementary weak-backscattering (quasiballistic) limit in Ref.~\cite{KBM}, which both found a mapping to a boundary sine-Gordon model with a free $\mathbb{Z}_3$ parafermion at its critical point.

Opening the second outer QPC on each island realizes the full charge-Kondo quartet setup considered in this Letter. This allows us to explore the fate of two coupled 2CK critical states: does inter-island coupling preserve local criticality, generate collective NFL behavior, or restore Fermi-liquid screening? This addresses a broader question concerning the stability of local NFL states to intersite coupling, relevant also to the validity of dilute-limit descriptions of multichannel impurity systems~\cite{Cox_Adv_Phys(47)_1998}. Such circuit simulators may also provide a controlled starting point for exploring the interplay between local criticality and non-local coupling in a lattice. A key challenge is whether boundary-critical impurity dynamics can survive in a self-consistent lattice environment and support a genuine bulk NFL phase, despite competing coherence and ordering tendencies.

Our results show that coupled critical states \textit{can} sustain correlated NFL regimes or give way to Fermi-liquid screening, depending on the interaction strength.

\clearpage
\textbf{\emph{Quantum circuit model.--}} 
The physical circuit considered here, following the experimental setup of Ref.~\cite{pouse2021}, is illustrated in Fig.~\ref{fig1}. It consists of two identical metallic islands or grains ($\alpha=L,R$) with large charging energies $E_C$ and effectively vanishing level spacings,  in a spin-polarized quantum Hall regime. Each island is tunnel-coupled to two external leads through tunable single-channel quantum point contacts (QPCs). The two islands are coupled together at a central QPC. The microscopic Hamiltonian can be decomposed as $H= H_0+H_c+H_{\rm QPC}$, where,
\begin{eqnarray}
&H_0 = \sum_{j=1}^5\; \sum_{\gamma=1}^2\;\sum_k \epsilon_k^{\phantom\dagger} \psi_{j\gamma k}^\dagger\psi_{j\gamma k}^{\phantom\dagger} ,\label{eq:H0}\\
&H_{QPC} = \sum_{j=1}^5 \left [ t_{j}^{\phantom\dagger} \psi_{j 1}^\dagger(0)\psi_{j2}^{\phantom\dagger}(0)  + {\rm H.c.}\right ], \label{eq:Hqpc}\\
&H_c = E_C\sum_{\alpha=L,R} \left(\hat N_\alpha-N_{\alpha}^0-N_{g\alpha}\right)^2 ,\label{eq:Hc}
\end{eqnarray}
where $\psi_{j\gamma k}^{\dagger}$ creates an electron of energy $\epsilon_k$ in continuum $\gamma$ either side of QPC $j$, while $\psi_{j\gamma}^\dagger(0)$ is the corresponding local operator at the contact. Tunneling across QPC $j$ is controlled by $t_j$. 
We take $j=1,2$ ($3,4$) to be lead contacts attached to the left (right) island; and $j=5\equiv C$ to be the central QPC. 
Here $\gamma=1$ (2) denotes lead (island) electrons for $j=1...4$.
For the central QPC we take $\gamma=1\equiv L$ or $\gamma=2\equiv R$ to be electrons on the $L$ or $R$ islands.
The total charge on the islands is then $\hat{N}_{L}=\sum_k \left[ \sum_{j=1}^2 \psi_{j 2 k}^\dagger\psi_{j 2 k}^{\phantom\dagger}+ \psi_{C L k}^\dagger\psi_{C L k}^{\phantom\dagger}\right ]$ and $\hat{N}_{R}=\sum_k \left[ \sum_{j=3}^4 \psi_{j 2 k}^\dagger\psi_{j 2 k}^{\phantom\dagger}+ \psi_{C R k}^\dagger\psi_{C R k}^{\phantom\dagger}\right ]$.  The excess filling of island $\alpha$ relative to $N_{\alpha}^0$ is controlled by the dimensionless parameter $N_{g\alpha}$, tuned by gate voltage $V_{g\alpha}$. The inter-island capacitive interaction is omitted; it can be restored without qualitatively changing the discussion. 
The QPCs are characterized experimentally by their bare transmissions $\tau_j$ (or equivalently the reflection amplitudes $|r_j|^2=1-\tau_j$), where $\tau_j=4g_j^2/(1+g_j^2)^2$ and $g_j=\pi\rho_0|t_j|$ with $\rho_0$ the Fermi level density of states.

Pinching off the central QPC gives two decoupled copies of the charge-2CK setup realized in  Refs.~\cite{Pierre_2015,Pierre_2018,piquard2026experimental}. On the other hand, coupling only one lead to each island gives the DCK model realized in Ref.~\cite{pouse2021}. Our full setup allows the coupling between 2CK states to be studied in a controlled setting. 
In the following we assume equal island-lead couplings $t_j\equiv t$ for $j=1...4$ and tune the central QPC coupling $t_5\equiv t_C$. 

For small-to-intermediate QPC transmissions and $T\ll E_C$ 
only the lowest two charge states near $N_{\alpha}^0$ contribute for each island, and these can be mapped to pseudospin-$\tfrac{1}{2}$ degrees of freedom. Tunneling at the QPCs causes pseudospin flips and generate Kondo exchange interactions~\cite{Furusaki1995b}. We treat this Kondo limit using Wilson's numerical renormalization group (NRG)~\cite{qmp3,qmp1,fdmnrg,mitchell2014generalized,stadler2016interleaved,weichselbaum2012non}. In the quasiballistic limit of strong QPC transmission (weak backscattering $|r_j|\ll 1$) we analyze the problem analytically using bosonization~\cite{Gogolin_book,delft} and quantum Brownian-motion~\cite{Yi,Yi1,qba} techniques.


\textbf{\emph{Kondo limit.--}} 
With $N_\alpha^0$ and $N_\alpha^0+1$ the two retained charge states on each island, we define $|\Downarrow\rangle_\alpha=|N_\alpha^0\rangle$ and $|\Uparrow\rangle_\alpha=|N_\alpha^0+1\rangle$ such that $S_\alpha^+=|\Uparrow\rangle_\alpha\langle\Downarrow|$ and $S_\alpha^z=\tfrac{1}{2}[|\Uparrow\rangle_\alpha\langle\Uparrow|
-|\Downarrow\rangle_\alpha\langle\Downarrow|]$.
Thus $\hat N_\alpha=N_\alpha^0+\tfrac12+S_\alpha^z$ in the projected
space, and Eq.~\ref{eq:Hc} becomes,
\begin{eqnarray}\label{eq:Hc_nrg}
H_c&= \sum_{\alpha} h_\alpha S_\alpha^z \quad ;\quad h_\alpha=2E_C\left(\tfrac12-N_{g\alpha}\right).
\end{eqnarray}
Following Refs.~\cite{Furusaki1995b,Sela_2016,Pierre_2018,pouse2021}, projecting Eq.~\ref{eq:Hqpc} yields,
\begin{eqnarray}
H_{QPC}=&t\left[S_{L}^+\left ( s^-_{1} + s^-_{2} \right ) + S_{R}^+\left ( s^-_{3} + s^-_{4} \right ) +{\rm H.c.}\right] \nonumber\\
&+t_C\left[S_L^+S_R^-\psi_{C L}^\dagger(0)\psi_{C R}^{\phantom\dagger}(0) +{\rm H.c.}\right]\;,  \qquad\label{eq:HK}
\end{eqnarray}
where $s^-_{j}=\psi_{j2}^\dagger(0)\psi_{j1}^{\phantom\dagger}(0)$ for $j=1...4$. $H_0$ remains as in Eq.~\ref{eq:H0}; we take the conduction electron density of states to be flat with a half-bandwidth $D\equiv 1$. 

The effective model is a generalization of two-channel-two-impurity (`Kondo quartet') system~\cite{qqt} but with $t_C$ embodying correlated inter-impurity charge transfer. 

We solve this model using NRG at zero field $h_L=h_R=0$, corresponding to the point of island charge degeneracy $N_{gL}=N_{gR}=\tfrac{1}{2}$. With ten spinless continua, the system is at the limit of what can be tackled numerically, even when leveraging the interleaved `iNRG' method~\cite{mitchell2014generalized,stadler2016interleaved} and exploiting full non-Abelian symmetries~\cite{weichselbaum2012non}. Details are provided in the \textit{End Matter}. 

\begin{figure*}[t]
    \centering
    \begin{minipage}[t]{0.7\textwidth}
        \vspace{0pt}
        \includegraphics[width=\linewidth]{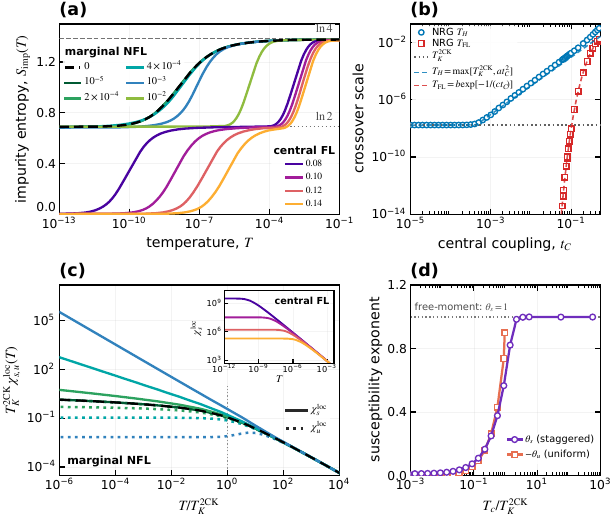}
    \end{minipage}
    \hfill 
    \begin{minipage}[t]{0.28\textwidth}
\caption{NRG results for the two-island circuit in the Kondo limit. We set the island-lead coupling to $t=0.1$  throughout and vary the inter-island coupling $t_C$. Panel~(a) shows the impurity entropy $S_{\rm imp}(T)$ for various $t_C$ as indicated in the legend. Panel (b) shows the extracted  scales $T_H$ and $T_{FL}$. The dotted line shows the pure 2CK scale $T_K^{2CK}$ in the decoupled limit $t_C=0$, while dashed lines show functional forms at finite $t_C$. For $T_H\gg T_K^{2CK}$ we see the emergence of a locking scale $T_c\sim t_C^2/D$ whereas the FL scale is of Kondo form, $T_{FL}\sim D e^{-1/ct_C}$.
Panel~(c) shows the behavior of the local staggered and uniform charge susceptibilities $\chi_{s,u}^{\rm loc}(T)$ for the same systems as (a). Susceptibility power-law exponents plotted in panel (d)  extracted from fits to:}
        \label{fig:NRG}
\vspace{6pt}
{\small
\begin{equation}
\chi_{y}^{\mathrm{loc}}(T) =  b_y+a_y
 \frac{
 \left(T_K^{\rm 2CK}/T\right)^{\theta_y}-1
 }{\theta_y}.
 \label{eq:chi}
\end{equation}
}
    \end{minipage}
\end{figure*}

A summary of our numerical results is presented in Fig.~\ref{fig:NRG}. We focus here on the thermodynamic impurity entropy $S_{\rm imp}(T)$ and static uniform/staggered local impurity susceptibilities $\chi_{u,s}^{\rm loc}(T)$. From these we extract Kondo scales and critical exponents. 
Our results show that in the symmetric Kondo limit, a weak inter-island coupling does \textit{not} destabilize the individual 2CK critical points of the left and right islands. As the coupling increases,  
the islands enter a correlated crossover regime with continuously varying susceptibility exponents, consistent with a marginal line of NFL fixed points. 
Eventually, at much larger $t_C\sim t$, we observe a quantum phase transition to a screened FL state, driven by an inter-island charge-Kondo effect at the central QPC.

In Fig.~\ref{fig:NRG}(a) we plot NRG results for the impurity entropy $S_{\rm imp}$ of the island pseudospins upon decreasing the temperature $T$, for fixed island-lead coupling $t$. The different lines correspond to different inter-island couplings $t_C$, as indicated in the legend. The black dashed line ($t_C=0$) shows the exactly-decoupled limit of two independent charge-2CK models, with a crossover from $S_{\rm imp}=2\ln 2=\ln 4$ at high-$T$ for two free island pseudospins, to $S_{\rm imp}=2\ln \sqrt{2}=\ln 2$ below a critical 2CK scale $T_K^{2CK}$ corresponding to two free Majoranas~\cite{EK1}. At finite $t_C$ we identify a scale $T_H$ characterizing the entropic crossover $S_{\rm imp}=\ln 4 \to \ln 2$ (defined in practice by $S_{\rm imp}(T_H)=\tfrac{3}{2}\ln 2$). As shown by the blue points in Fig.~\ref{fig:NRG}(b), $T_H\simeq T_K^{2CK}$ for $T_c \ll T_K^{2CK}$, where $T_c \sim t_C^2/D$ is the perturbative `locking' scale which lowers the energy of island states $|\Uparrow,\Downarrow\rangle$ and $|\Downarrow,\Uparrow\rangle$ relative to $|\Uparrow,\Uparrow\rangle$ and $|\Downarrow,\Downarrow\rangle$. In this regime of small finite inter-island coupling $t_C$ (dark green lines), we find that the residual $T=0$ entropy remains pinned to $S_{\rm imp}=\ln 2$. For these same systems, panel (c) shows the local uniform and staggered island pseudospin susceptibilities, which are equal and follow the pure 2CK result, $\chi_u^{\rm loc}=\chi_s^{\rm loc}\sim \ln(T_K^{2CK}/T)$. This establishes essentially independent 2CK effects for $T_c\ll T_K^{2CK}$.

For $T_c\sim T_K^{2CK}$ we see a rapid crossover behavior in the susceptibilities (Fig.~\ref{fig:NRG}c). Solid lines show the staggered susceptibility diverging as a power-law upon reducing temperature, $\chi_s^{\rm loc}\sim (T_K^{2CK}/T)^{\theta_s}$. Dotted lines for the uniform susceptibility approach a constant as $\chi_u^{\rm loc}(T)-\chi_u^{\rm loc}(0)\sim (T_K^{2CK}/T)^{\theta_u}$. Here $0\le \theta_s\le 1$ but $\theta_u \le 0$. 
Eq.~\ref{eq:chi} continuously interpolates the staggered response $\chi_s^{\rm loc}(T)$ between the logarithmically divergent 2CK behavior for $\theta_s\to 0$ and Curie behavior $\sim 1/T$ as $\theta_s\to 1$. The latter limit indicates a \textit{free} (unscreened) local moment comprising the degenerate pair of island states  $|\Uparrow,\Downarrow\rangle$ and $|\Downarrow,\Uparrow\rangle$. The uniform susceptibility saturates at low-$T$, giving a Van Vleck type response, indicating that we have a gapped uniform-charge sector. 
We plot the numerically-extracted exponents in Fig.~\ref{fig:NRG}(d). The intermediate exponents characterize a marginal line of NFL fixed points.

As seen from the entropy curves in Fig.~\ref{fig:NRG}(a),
a new low-energy crossover $S_{\rm imp}=\ln 2 \to 0$ emerges
for much larger $t_C\sim t$. This corresponds to the Kondo screening of the inter-island local moment by electronic tunneling at the central QPC which interconverts $|\Uparrow,\Downarrow\rangle \leftrightarrow |\Downarrow,\Uparrow\rangle$, and results in a FL state for $T\ll T_{FL}$. Here $T_{FL}$ is defined via $S_{\rm imp}(T_{FL})=\tfrac{1}{2}\ln 2$, and is plotted as the red points in Fig.~\ref{fig:NRG}(b). It follows the standard single-channel Kondo form~\cite{Hewson}. The staggered charge dynamics are therefore Kondo-quenched below $T_{FL}$, as shown in the inset to panel (c).


\textbf{\emph{Quasiballistic limit.--}} 
To uncover the full phase diagram and to get a detailed analytic understanding of the problem, we turn now to the strong transmission regime, where couplings are most naturally described in terms of the (small) reflection amplitudes $|r|$ and $|r_C|$.

Here we can employ the bosonization framework~\cite{bon1,WE1, Kane_1992}, extending the analysis of Ref.~\cite{KBM} for the DCK model to the present setup. We write Eq.~\ref{eq:H0} as,
\begin{eqnarray}
H_{\rm 0}=\frac{v_{\rm F}}{4\pi}\sum_{j=1}^5\int^\infty_{-\infty} dx\left[\left(\partial_x\varphi_{j, l}\right)^2+\left(\partial_x\varphi_{j, r}\right)^2\right],\label{eq1}
\end{eqnarray}
where  $v_{\rm F}$ denotes the Fermi velocity, and the bosonic field $\varphi_{j,l/r}(x,t)$ corresponds to the left- and right-moving chiral fermions which satisfy the commutation relations $\left[\varphi_{j,l/r}(t),\varphi_{j,l/r}(t_1)\right] = \pm i\pi\,\mathrm{sgn}(t-t_1)$.

For the QPCs described by Eq.~\ref{eq:Hqpc}, a weak-backscattering expansion yields a strongly nonlinear boundary Hamiltonian~\cite{Kane_1992}. With compact bosonic fields defined as $\delta\varphi_j=\varphi_{j,r}-\varphi_{j,l}$ we can write, 
\begin{eqnarray}
H_{QPC}=-\sum_{j=1}^5\frac{D|r_{j}|}{\pi}\cos \left[\delta\varphi_j(0, t)\right] \;,
\end{eqnarray}
where $D$ is the high-energy cutoff.

The charging Hamiltonian Eq.~\ref{eq:Hc} similarly follows, using the island number operators,
\begin{align}
\hat{N}_L= &\int^\infty_0 \!\frac{dx}{2\pi}\left [\partial_x\delta\varphi_1(x)+\partial_x\delta\varphi_2(x)\right ] {+}\int^0_{-\infty}\! \frac{dx}{2\pi} \partial_x\delta\varphi_5(x),\nonumber\\
\hat{N}_R= &\int^0_{-\infty} \!\frac{dx}{2\pi}\left [\partial_x\delta\varphi_3(x)+\partial_x\delta\varphi_4(x)\right ] {+}\int_0^{\infty}\! \frac{dx}{2\pi} \partial_x\delta\varphi_5(x).\nonumber
\end{align}


\textbf{\emph{Effective action.--}}
We define spin (s) and charge (c) field combinations on the left (L) and right (R) sides,  
$\Phi_{Ls}=(\delta\varphi_1-\delta\varphi_2)/\sqrt{2}$,~~ $\Phi_{Lc}=(\delta\varphi_1+\delta\varphi_2)/\sqrt{2}$~~and $\Phi_{Rs}=(\delta\varphi_3-\delta\varphi_4)/\sqrt{2}$,~~ $\Phi_{Rc}=(\delta\varphi_3+\delta\varphi_4)/\sqrt{2}$.~~We then transform the charge combinations according to,
\begin{eqnarray}
\begin{pmatrix}
\Phi_a\\[2pt]
\Phi_b\\[2pt]
\Phi_c
\end{pmatrix}
=
\begin{pmatrix}
-\tfrac{1}{2} & -\tfrac{1}{2} & \tfrac{1}{\sqrt{2}}\\[2pt]
-\tfrac{1}{\sqrt{2}} & \tfrac{1}{\sqrt{2}} & 0\\[2pt]
\tfrac{1}{2} & \tfrac{1}{2} & \tfrac{1}{\sqrt{2}}
\end{pmatrix}
\begin{pmatrix}
\Phi_{Lc}\\[2pt]
\Phi_{Rc}\\[2pt]
\delta\varphi_5
\end{pmatrix}
\label{eq:bosonic_rotation}
\end{eqnarray}
such that the island charges $\hat{N}_L$ and $\hat{N}_R$ depend only on $\Phi_a$ and $\Phi_b$. The charging term Eq.~\ref{eq:Hc} consequently gaps these modes, which allows us to integrate them out at low temperatures $T\ll E_C$. We then obtain an effective action of the form $S_{\rm eff}=S_0+\int^\beta_0d\tau\; \mathcal{H}_{\rm b}(\tau)$ with
\begin{eqnarray}
S_0&=\frac{1}{2\pi\beta}\sum_n|\omega_n|\left(|\Phi_{Ls}|^2+|\Phi_{Rs}|^2+|\Phi_{c}|^2\right),
\end{eqnarray}
where $\beta=1/T$ and $\omega_n=2\pi n T$ are the  Matsubara frequencies. In the L/R-symmetric case $N_{gL}=N_{gR}\equiv N_g$, the effective boundary Hamiltonian becomes,
\begin{equation}
\mathcal{H}_{\rm b} =-\sum_{\alpha}\lambda \cos\Phi_{ \alpha s}\cos\!\left(\frac{\Phi_c}{2}-\sigma_\alpha\pi N_g\right) -\lambda_C \cos\!\Phi_c, \label{eqbh}
\end{equation}
with $\sigma_\alpha=\pm 1$ for $\alpha=L,R$. The couplings are given by,
\begin{eqnarray}
\lambda= \tfrac{\sqrt8}{\pi} \zeta^{3/8}D\:|r| ~~~ ; ~~ \lambda_C=\tfrac{2}{\pi}\zeta^{1/2}D\:|r_C|,
\end{eqnarray}
where $\zeta=e^{\gamma} E_C/\pi D$ and $\gamma$ is Euler's constant.

In the basis of $(\Phi_{Ls},\Phi_{Rs},\Phi_c)$, the backscattering Hamiltonian Eq.~\ref{eqbh} is periodic under the translations
$\mathsf T_1=(2\pi,0,0)$, $\mathsf T_2=(0,2\pi,0)$, and
$\mathsf T_3=(\pi,\pi,2\pi)$. These vectors generate the direct lattice of the quantum Brownian motion (QBM) problem~\cite{Yi, Yi1, qba}. The reciprocal vectors, defined by $\mathsf G_i\!\cdot\!\mathsf T_j=2\pi\delta_{ij}$
are given by  $\mathbf{G}_{1}=(1,0,-1/2)$, $\mathbf{G}_{2}=(0,1,-1/2)$, and $\mathbf{G}_{3}=(0,0,1)$. The norms of these vectors determine the scaling dimensions of the corresponding boundary operators. Since $|\mathbf{G}_{1}|^2=|\mathbf{G}_{2}|^2=5/4$ and $|\mathbf{G}_3|^2=1$, the operators associated with $\mathbf{G}_{1}$ and $\mathbf{G}_{2}$ have scaling dimension $\Delta_{L,R}=5/8$, while the operator associated with $\mathbf{G}_3$ has scaling dimension $\Delta_C=1/2$. All three perturbations are \textit{relevant} in the renormalization-group (RG) sense. 
Thus the couplings $\lambda$ and $\lambda_C$ grow under RG, and the initially small bare reflection amplitudes $|r|$ and $|r_C|$ become large as the temperature is lowered, as the system enters the strong-coupling regime.

Since the phase diagram is controlled by the competition between pinning at the outer barriers ($\lambda$) and pinning at the central barrier ($\lambda_C$), we must determine which perturbation reaches the strong barrier limit first. To do this, we introduce the dimensionless couplings $\mathcal{Y}(D)=\lambda/D$ and $\mathcal{Y}_c(D)=\lambda_C/D$. At tree level, the RG equations are $d\mathcal{Y}/d\ell=3\mathcal{Y}/8$ and $d\mathcal{Y}_c/d\ell=\mathcal{Y}_c/2$ with the running energy scale $\epsilon=De^{-\ell}$. Integrating these equations gives $\mathcal{Y}(\epsilon)=\mathcal{Y}(D)(D/\epsilon)^{3/8}$ and $\mathcal{Y}_c(\epsilon)=\mathcal{Y}_c(D)(D/\epsilon)^{1/2}$. The crossover scales $\epsilon^*\sim |r|^{8/3}E_C$ and $\epsilon_c^*\sim |r_C|^2E_C$ are identified when the running couplings become of order 1. The central QPC is thus the dominant strong barrier when $\epsilon_c^* \gg \epsilon^*$, or equivalently
$|r| \ll |r_C|^{3/4}$. This case, as well as the limit of a fully-ballistic central QPC, is considered in the \textit{End Matter}.

\begin{figure}[t]
\centering
\includegraphics[width=0.9\columnwidth]{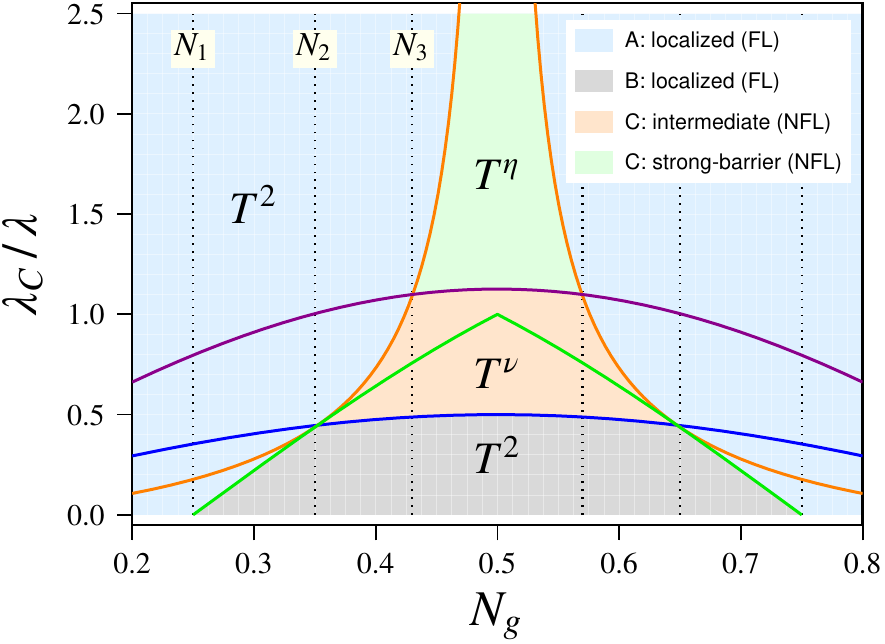}
\caption{Ground state phase diagram and scaling behavior of transport properties obtained by QBM. The blue, orange, green and purple curves denote $\mathcal{K}_{BC}$, $\mathcal{K}_{AC}$, $\mathcal{K}_{AB}$, and  $\mathcal{K}_{\rm st}$.}
\label{instanton}
\end{figure}


\textbf{\emph{Phase diagram.--}}
Since all three backscattering operators are relevant, their competition is determined by the minima of Eq.~\ref{eqbh}.  We set
$a=\pi N_g$, $w=\Phi_c/2$, and
$\mathcal K=\lambda_C/\lambda$~\footnote{Here $\mathcal K$ should be taken at a common strong-coupling matching scale.  At the bare cutoff we have $\mathcal K_0=2^{-1/2}(e^\gamma E_C/\pi D)^{1/8}|r_C|/|r|$, while the
weak-backscattering flow gives
$\mathcal K(E)=\mathcal K_0(D/E)^{1/8}$ at running energy $E$.}.  Minimizing independently over $\Phi_{Ls}$ and $\Phi_{Rs}$ reduces the problem to an effective 1d potential,
\begin{equation}
 \frac{V(w)}{\lambda}=-|\cos(w-a)|-|\cos(w+a)|
 -\mathcal K\cos 2w .
 \label{eq:Vmin}
\end{equation}
Charge periodicity and particle--hole symmetry allow us to restrict to
$0\leq N_g\leq 1/2$ and $0\leq w\leq\pi/2$; the other half of the
diagram follows under $N_g\mapsto1-N_g$.  There are three competing
minima, denoted A, B, and C:
\begin{align}
 w_A&=0,                 & V_A/\lambda&=-2\cos a-\mathcal K,
 \nonumber\\
 w_B&=\frac{\pi}{2},     & V_B/\lambda&=-2\sin a+\mathcal K,
 \nonumber\\
 \sin w_C&=\frac{\sin a}{2\mathcal K},
 &V_C/\lambda&=-\mathcal K-\frac{\sin^2a}{2\mathcal K}.
 \label{eq:three_minima}
\end{align}
The interior solution C exists for $\sin a\leq2\mathcal K\leq\tan a$. 
To obtain the phase boundaries plotted in Fig.~\ref{instanton}, we equate the energies in Eq.~\ref{eq:three_minima}. This gives,
\begin{align}
 \mathcal K_{BC}&=\frac{\sin a}{2},&
 \mathcal K_{AC}&=\frac{\sin^2a}{4\cos a},\nonumber\\[-2pt]
 \mathcal K_{AB}&=\sin a-\cos a .&&
 \label{eq:phase_boundaries}
\end{align}
From this, we can identify three gate thresholds, marked as $N_1$, $N_2$ and $N_3$ in Fig.~\ref{instanton}. $\mathcal K_{AB}$ first becomes positive at $N_1=\tfrac{1}{4}$, and so A is the ground state for $N_g<N_1$. For $N_1<N_g<N_2$ there is a direct first-order B--A transition along $\mathcal K_{AB}$.  All three boundaries meet at
$N_2=\arctan(2)/\pi\simeq0.3524$. Therefore when $N_2<N_g<1/2$, C is the ground state for $\mathcal K_{BC}<\mathcal K<\mathcal K_{AC}$. This gives the basic structure of the phase diagram.

\textbf{\emph{QBM fixed points and transport.--}}
The strong-barrier structure is fixed by the lattice of minima of $\mathcal{H}_b$. The minima repeat under the translations $\mathsf T_{1,2,3}$ given
above. In phases A and B this lattice has a single-site basis, so the only instantons are the Bravais-lattice translations $\mathsf T_{1,2}$. 
Phase C instead splits into two inequivalent sublattices, with a two-site basis represented by
$\mathsf Q_+=(0,\pi,2w_C)$ and $\mathsf Q_-=(\pi,0,-2w_C)$. This generates an additional class of inter-sublattice instantons that controls the stability of phase C~\cite{Yi}.

For a displacement $\mathsf S$, the fugacity (or tunneling amplitude) $t_{\mathsf S}$ of an instanton is governed by the scaling equation $dt_{\mathsf S}/d\ell=(1-\Delta)t_{\mathsf S}$~\cite{Yi,Yi1,qba}. The instanton is therefore RG relevant for scaling dimensions $\Delta<1$. QBM theory gives $\Delta=|\mathsf S|^2/(2\pi^2)$, and 
so the corrections to conductance $\delta G \sim T^{2\Delta-2}$ are controlled by $\mathsf S$. 

In phases A and B the shortest instantons are
$\mathsf T_{1,2}$, so $\Delta_A=\Delta_B=2$ and
$\delta G(T)\sim T^2$. These are localized FL-like phases. 
In phase C the two shortest inequivalent inter-sublattice displacements and their dimensions are,
\begin{align}
 \mathsf S_1&=(\pi,-\pi,-4w_C),
 &\Delta_1&=1+8\left(\frac{w_C}{\pi}\right)^2,
 \nonumber\\
 \mathsf S_2&=(0,0,2\pi-4w_C),
 &\Delta_2&=2\left(1-\frac{2w_C}{\pi}\right)^2.
 \label{eq:instanton_dimensions}
\end{align}
Since $\Delta_1>1$ for $w_C>0$, strong-barrier stability is lost at $\Delta_2=1$. This gives an internal phase-C  boundary,
\begin{equation}
 w_{\rm st}=\frac{\pi}{2}\left(1-\frac{1}{\sqrt2}\right),\qquad
 \mathcal K_{\rm st}=\frac{\sin(\pi N_g)}{2\sin w_{\rm st}}.
 \label{eq:Kst}
\end{equation}
This line meets $\mathcal K_{AC}$ at
$N_3=\pi^{-1}\arctan[2/\sin w_{\rm st}]\simeq0.4305$.  Thus C is entirely an intermediate-coupling NFL for $N_2<N_g<N_3$; for
$N_3<N_g<1/2$, it is intermediate-coupling for
$\mathcal K_{BC}<\mathcal K<\mathcal K_{\rm st}$ and a stable strong-barrier NFL for $\mathcal K_{\rm st}<\mathcal K<\mathcal K_{AC}$.

In the strong-barrier NFL, the leading dimension is
$\Delta_s=\min(\Delta_1,\Delta_2)$: it equals $\Delta_1$ for
$w_C<\pi/8$ and $\Delta_2$ for $\pi/8<w_C<w_{\rm st}$.  Hence $1<\Delta_s\leq9/8$ and
\begin{equation}
 \delta G(T)\sim T^\eta,\qquad
 \eta=2(\Delta_s-1),\qquad 0<\eta\leq\frac14 .
 \label{eq:transport_exponents}
\end{equation}
By contrast, for $\mathcal K_{BC}<\mathcal K<\mathcal K_{\rm st}$ we have $0<\Delta_2<1$ so both the weak- and strong-barrier endpoints are unstable.
The two-sublattice QBM problem therefore flows to a stable fixed point with finite universal mobility $G^*$~\cite{Yi,Yi1}, with $G(T)-G^*\sim T^\nu$ and
$\nu>0$.  This intermediate-coupling exponent is not determined by the endpoint dimensions in Eq.~\ref{eq:instanton_dimensions}.


 \textbf{\emph{Discussion.--}} The two-site charge Kondo circuit realized experimentally in Ref.~\cite{pouse2021} offers a remarkable opportunity to study the fate of coupled NFL states in a controlled and tunable setting. Here we analyze the problem in the Kondo limit of weak/intermediate transmission, and in the opposite quasiballistic limit, obtaining complementary numerical and analytical insights into the underlying physics.

 We find that the separate critical 2CK states in the decoupled-island limit survive relatively weak tunneling at the central QPC, and continuously evolve into a NFL phase that is robust to moderate gate detuning. This phase is characterized by nontrivial transport and susceptibility exponents. Stronger coupling at the central QPC or strong gate detuning results in a transition to a regular FL phase. 
 These signatures should be experimentally accessible through gate-dependent conductance scaling and island-charge sensing.

 Our results may have implications for the stability of bulk NFL phases in lattices, and for the prospects of manipulating multiple Kondo anyons in artificial networks. \\


\begin{acknowledgments}
\textit{\textbf{Acknowledgments.--}}
We thank David Goldhaber-Gordon and Frederic Pierre for enlightening discussions about practical implementation. A.K.M.~acknowledges funding support from Taighde \'Eireann - Research Ireland under Grant No.~24/FFP-P/12816 and through Rinn Quantum under grant No.~23/RC/12197. 

\vfill
\clearpage


\appendix
\section*{End Matter}
\label{sec:endmatter}

\setcounter{figure}{0}
\renewcommand{\thefigure}{E\arabic{figure}}
\setcounter{equation}{0}
\renewcommand{\theequation}{E.\arabic{equation}}
\setcounter{table}{0}
\renewcommand{\thetable}{E.\arabic{table}}

\textbf{\emph{NRG implementation and observables.--}}
We use NRG~\cite{qmp3,qmp1} to solve the projected quantum impurity Hamiltonian $H=H_0+H_c+H_{\rm QPC}$ defined by Eqs.~\ref{eq:H0}, \ref{eq:Hc_nrg} and \ref{eq:HK}. We specialize to the channel-symmetric, charge-degenerate case, with $t_i\equiv t$ for $i=1...4$ and $h_L=h_R=0$, where the latter condition corresponds to $N_{gL}=N_{gR}=\tfrac{1}{2}$. Each of the five QPCs is represented by two spinless electronic continua, giving ten flat conduction bands of half-bandwidth $D\equiv1$, which sets the unit of energy. This Hamiltonian has the exact symmetry
$U(1)_S\times Sp(4)_L\times Sp(4)_R\times SU(2)_C$.
The non-Abelian $Sp(4)_\alpha$ symmetry combines channel rotations with axial-charge transformations of the two outer QPCs attached to island $\alpha=L,R$, while $SU(2)_C$ is the axial-charge symmetry of the two continua forming the central QPC. The Kondo exchange is purely transverse, leaving a total $U(1)_S$ pseudospin symmetry. We formulate the iterative NRG diagonalization directly in a symmetry-adapted multiplet basis, using reduced matrix elements of operators and Clebsch--Gordan coefficients~\cite{weichselbaum2012non}. At each complete Wilson shell, the ten spinless Wilson orbitals span $2^{10}=1024$ local states, but the non-Abelian symmetry decomposition organizes these into only $108$ reduced local multiplets, substantially reducing the computational cost.

However, even leveraging full symmetries, the effective 5-channel problem is still intractable in practice with standard NRG. We therefore additionally use the `interleaved' NRG (iNRG) method~\cite{mitchell2014generalized,stadler2016interleaved}
which breaks each complete Wilson iteration into several substeps. By interleaving different Wilson chains with modified discretizations into a single generalized Wilson chain~\cite{mitchell2014generalized}, additional truncations can be performed between substeps, dramatically reducing the dimensions of the multiplet blocks to be diagonalized. Only by exploiting full symmetries and iNRG was a numerical solution of the two-site charge-Kondo model feasible.

In practice we used NRG discretization parameter $\Lambda=6$ with an energy-based truncation scheme, which resulted in roughly 1--4 million states being retained at each iNRG substep. We verified that the reported physical quantities were accurately converged with respect to the number of kept states and $\Lambda$.

The thermodynamic entropy was computed directly using the complete Anders-Schiller basis across all iNRG substeps~\cite{anders2006spin}. We report the impurity contribution $S_{\rm imp}(T)=S_{\rm full}(T)-S_{\rm bath}(T)$ where $S_{\rm bath}$ was obtained from an independent iNRG run with the same parameters but without the island pseudospins. The results of two different $z$-shift calculations were averaged~\cite{vzitko2009energy}. For display purposes (only) the data were further smoothed in post-processing to remove discretization artifacts. 

Static local susceptibilities were obtained from the full density matrix~\cite{fdmnrg}. We define the island-charge operators and their local susceptibilities as,
\begin{align}
 Q_u&=S_L^z+S_R^z, & Q_s&=S_L^z-S_R^z, \nonumber\\[-2pt]
 \chi_{\nu}^{\rm loc}(T)
 &=\int_0^{1/T}\!d\tau\,
 \big\langle\delta Q_\nu(\tau)\delta Q_\nu(0)\big\rangle,
 &\nu&=u,s,
 \label{eq:EM_chidef}
\end{align}
with $\delta Q_\nu=Q_\nu-\langle Q_\nu\rangle$.  Equivalently,
$\chi_\nu^{\rm loc}=\partial\langle Q_\nu\rangle/\partial B_\nu|_{B_\nu=0}$ for $H_B=-B_uQ_u-B_sQ_s$. In practice we evaluate $\chi_\nu^{\rm loc}(T)$ via its Lehmann representation~\cite{fdmnrg}.

With $t=0.1$ and $t_C=0$ we found 
$T_K^{\rm 2CK}\simeq 10^{-8}D$.  We define the scales plotted in Fig.~\ref{fig:NRG}(b) by,
\begin{align}
 S_{\rm imp}(T_H)&=\tfrac32\ln2,
 &S_{\rm imp}(T_{\rm FL})&=\tfrac12\ln2,
 \label{eq:EM_scales_def}
\end{align}
with $T_K^{\rm 2CK}=T_H$ at $t_C=0$. The high-energy scale is,
\begin{align}
 T_H&\simeq\max(T_K^{\rm 2CK},T_c)\quad;\quad T_c\simeq at_C^2/D,
 \label{eq:EM_Tcfit}
\end{align}
with $a\simeq 0.1$. For the low-energy scale, we find
\begin{eqnarray}
 T_{\rm FL}/D=b\exp\!\left(-1/ct_C\right),
 \label{eq:EM_TFLfit}
\end{eqnarray}
where $b\simeq 0.74$ and $c\simeq 1.1 \rho_0$. 
The fitted value of $c$ is close to the pure transverse-exchange result for single-channel Kondo of \(4\rho_0/\pi\simeq 1.27 \rho_0\). The discrepancy can be attributed to mild outer-channel renormalization effects. 
We comment here that our NRG data alone cannot resolve whether $T_{FL}$ actually vanishes at finite $t_C$ or just becomes exceedingly small. 

The \texttt{Dub-NRG} codebase was developed in-house. It was refactored using GPT-5.6 Sol which also implemented non-Abelian symmetries. Benchmark outputs were validated against human-written Abelian code.\\


\textbf{\emph{Details of the QBM construction.--}}
Minimizing Eq.~\ref{eqbh} over $\Phi_{Ls}$ and $\Phi_{Rs}$ gives Eq.~\ref{eq:Vmin}.  For $a=\pi N_g$, $0\leq a,w\leq\pi/2$, its absolute values may be resolved,
\begin{equation}
 \frac{V(w)}{\lambda}=
 \begin{cases}
 -2\cos a\cos w-\mathcal K\cos2w,
       &0\leq w\leq\frac{\pi}{2}-a,\\
 -2\sin a\sin w-\mathcal K\cos2w,
       &\frac{\pi}{2}-a\leq w\leq\frac{\pi}{2}.
 \end{cases}
 \label{eq:EM_piecewiseV}
\end{equation}
The endpoint extrema give A and B, while stationarity of the second branch gives $\sin w_C=\sin a/(2\mathcal K)$.  Requiring $w_C$ to lie on that branch yields $\sin a\leq2\mathcal K\leq\tan a$; comparison of the three energies then gives Eq.~\ref{eq:phase_boundaries} and the thresholds $N_{1,2}$ quoted in the main text.

At strong backscattering, tunneling between neighboring minima is represented by instantons of displacement $\mathsf S$. Their corresponding scaling dimensions are given by $\Delta=|\mathsf S|^2/2\pi^2$. 
The one-site A and B lattices have shortest displacement $|\mathsf S|=2\pi$, hence $\Delta=2$.  For C, the two inequivalent inter-sublattice displacements give Eq.~\ref{eq:instanton_dimensions}; $\Delta_2=1$ produces $\mathcal K_{\rm st}$, and comparison with $\mathcal K_{AC}$ gives $N_3$.  When $\Delta_2<1$, both endpoint theories are unstable and the nonsymmorphic two-sublattice QBM problem flows to the finite-mobility intermediate fixed point~\cite{Yi,Yi1}.

\textbf{\emph{Strongly backscattering central QPC.--}} 
One can treat a strong central barrier as a weak perturbation to the two quasiballistic charge-2CK islands~\cite{mkk}. However, even within the fully quasiballistic limit, one can still consider the regime of strong \textit{renormalized} backscattering at the central QPC. This arises at low energies when $\epsilon_m^* \gg \epsilon_{l,r}^*$, or equivalently
$|r| \ll |r_C|^{3/4}$. 

In this limit, the perturbation at the central QPC is the largest energy scale, so we integrate out the charge mode $\Phi_c$ first. The corresponding cosine term pins $\Phi_c$ near one of its minima, while fluctuations about that minimum renormalize the scaling dimensions of the backscattering operators at the left and right QPCs.
Choosing one minimum, we write
\begin{equation}
\Phi_c(\tau)=2\pi n+\widetilde{\Phi}_c(\tau), \qquad n\in\mathbb Z .
\end{equation}
The low-energy fluctuations are then described by the quadratic action
\begin{equation}
S_{0}^{\rm g}
=\frac{1}{2\pi\beta}\sum_n \bigl(|\omega_n|+\mathscr{M}\bigr)\,
\widetilde{\Phi}_c(i\omega_n)\widetilde{\Phi}_c(-i\omega_n),
\end{equation}
where $\mathscr{M}$ is the renormalized pinning scale of the fluctuation field
and is of order $\epsilon_m^*$. Integrating out $\widetilde{\Phi}_c$ gives,
to leading order, the effective action
\begin{equation}
S_{\rm eff}^{(1)}
=-\lambda\mathcal{F}_1(\mathscr{M})\cos(\pi N_g)
\sum_{\alpha=L,R}
\int_0^\beta d\tau\,\cos\Phi_{\alpha s}(\tau),\label{ado1}
\end{equation}
with $\mathcal{F}_1(\mathscr{M})
=\left(e^{\gamma}\mathscr{M}/D\right)^{1/8}$.
At charge degeneracy, the leading-order theory reduces to two decoupled quasiballistic 2CK models~\cite{Furusaki1995b}. Physically, in the limit of an infinitely strong central barrier, the device is cut into two independent subsystems, each of which exhibits 2CK physics. At second order, pinning of the central QPC field generates a tree-level marginal operator corresponding to charge-transfer between the two islands,
\begin{equation}
S_{\rm eff}^{(2)}
\sim
-\lambda^2\,\mathcal{F}_2(\mathscr{M})
\sum_{\sigma=\pm} \int_0^{\beta}
d\tau \cos\!\left[\Phi_{Rs}(\tau)+\sigma\Phi_{Ls}(\tau)\right] + ...\label{ado2}
\end{equation}
where $\mathcal{F}_2$ is an analytic function of $\mathscr{M}$ and $N_g$, and $S_{\rm eff}^{(2)}$ remains finite at $N_g=1/2$.


\textbf{\emph{Fully open central QPC.--}} 
For $\lambda_C=0$ the central QPC is perfectly transmitting ($\tau_C=1$). In this case, the charges on the two islands cease to be quantized separately and only their sum remains quantized~\cite{dde1}. 
The collective field $\Phi_c$ thus remains dynamical and is not pinned. The outer operators retain dimension $5/8$ and are RG relevant; Eq.~\ref{eq:EM_piecewiseV} at $\mathcal K=0$ therefore selects A for $N_g<N_1$ and B for $N_g>N_1$. 

However, at \(N_g=N_1=1/4\) the A and B minima are degenerate and a 4CK-related critical problem can arise~\cite{Nozieres_Blandin_JPhys_1980,Yi1}. Because the collective-charge kernel and charge constraints differ from those of the canonical 4CK model, this special point is best viewed as an environmentally dressed quasiballistic 4CK realization.


\end{acknowledgments}

\end{document}